\documentclass[journal]{paper}

\usepackage{color,array}

\usepackage{graphicx}
\usepackage{caption}
\usepackage{csvsimple}
\usepackage{amsmath}
\usepackage{mathtools}

\begin{document}

\title{Domain Specific Concept Drift Detectors for Predicting Financial Time Series}

\author{Filippo Neri \\ University of Naples, ITALY \\ filippo.neri.email@gmail.com}

\maketitle

\begin{abstract}
Concept drift detectors are used in combination with learning systems to maintain a good accuracy rate on non-stationary data streams\footnote{Concept drift detection is a fundamental step in the concept drift adaptation process employed in learning systems to deal with dynamically varying concepts in data streams.}. Financial time series are an instance of non-stationary data streams whose concept drifts (market phases) are so important to affect investment decisions worldwide. This paper studies how concept drift detectors behave when applied to financial time series. General results are: a) concept drift detectors usually improve the runtime over continuous learning, b) their computational cost is usually a fraction of the learning and prediction steps of even basic learners, 
c)  it is important to study concept drift detectors in combination with the learning systems they will operate with, and d) concept drift detectors can be directly applied to the time series of raw financial data and not only to the model's accuracy one.
Moreover, the study introduces three simple concept drift detectors, tailored to financial time series, and shows that two of them can be at least as effective as the most sophisticated ones from the state of the art when applied to financial time series.
\end{abstract}


Keywords: Artificial intelligence, Learning systems, Concept drift detection, Financial time series modeling.

\section{Introduction}
Concept drifts in a data stream happen when the statistical properties, 
 governing the data classification, vary. 
 An introduction to concept drift and a review of standard concept drift detection methods can be found in \cite{driftsurvey,driftsurvey1}. 
Instances of potentially infinite and non-stationary data streams are those generated by IoT sensors, e-commerce applications, telecom networks, and financial trading systems. 
Concept drift detection has become an active research area together with the increased adoption of learning systems to deal with data streams.

The timely detection of concept drifts in a data stream allows for the prompt re-tuning of the model analyzing the data (adaptation to the concept drift) in order to maintain the best possible accuracy rate over time.
Most concept drift detectors in the literature are {\em general purpose} (domain indifferent) and usually monitor the model's accuracy rate (its error rate) over time to recognize when a concept drift happened in the data stream.  

Financial time series, an instance of non-stationary data streams, are commonly analyzed by the finance industry in order to make investment decisions. 
For financial companies, the timely detection of concept drifts (market phases) in financial time series would allow for the prompt re-learning of their analytical models and could lead to a more effective exploitation of learning systems.

This study investigates the application of concept drift detectors (c.d.d.) to financial time series by comparing general purpose c.d.d. and novel domain specific ones. 
The work support the thesis that c.d.d. cannot be studied as stand-alone algorithms, but should instead be analyzed taking into account the domain and the learners they are expected to work with.
 
The experimental study reported here shows that: \\
a) concept drift detectors reduce the resulting learning system's runtime with respect to using a continuous learning strategy that requires to learn a new model every time new data arrive. \\
b) the computational cost of using standard c.d.d. is small if compared to the learning and prediction steps of even basic learning algorithms. \\ 
c) it is beneficial to study c.d.d. in combination with the learning systems they will be combined with in order to understand the benefits and weaknesses of the resulting system. \\
d) c.d.d. can be directly applied to raw financial time series and not only to accuracy ones.
Feeding raw data from the time series to the c.d.d. is a novelty with respect to the state of the art to the best of our knowledge.\\
In addition, the study introduces three simple concept drift detectors (in terms of their algorithm) tailored to financial time series, and shows that two of them can be as effective as those  in the literature when applied to financial time series. The standard c.d.d. that we will consider are: EDDM \cite{eddm}, ADWIN \cite{adwin}, DDM \cite{ddm}, KSWIN \cite{kswin}, PH \cite{ph}, HDDM\_A \cite{hddm}, HDDM\_W \cite{hddm}. They will be studied in combination with the following standard 
regressors: Linear Regression \cite{linregr}, Bayesian Ridge Regressor \cite{brr}, and MLPR \cite{mlpr} plus two baseline ones, YC and MinValInTS, developed by the author. 
We leave for future research the study of how c.d.d. interact with complex learners such as
 deep learning \cite{deeptrading,deeptrading2}, technical analysis \cite{technicaltrading,techtrading2}, reinforcement learning \cite{reinforcementtrading,deeptrading2}, evolutionary computation \cite{NeriAIcom12} or agent based simulation \cite{NeriExSyst2020,Neri2019Wivace}.
 
 
The paper is organized as follow: Section \ref{stateofart} presents the state of the art on learning with concept drift detection, Section \ref{scdd} and \ref{sls} introduce the concept drift detectors and the learning algorithms used in the experiments. Section \ref{sliding} describes how sliding window learning can be combined with c.d.d. whereas Section \ref{efs} introduces a working hypothesis on the average prediction error on financial time series. The experimental part can be found in Section \ref{expSec}. 
Section \ref{compcost} comments on the computational cost of combining sliding window learning with
concept drift detection. Finally, the future work section concludes the paper. 

\section{State of the art \label{stateofart}}
Concept drift detectors are particularly useful when dealing with non-stationary data streams
as an alternative to continuous learning.
The latter, in fact, requires the update of the data model
after the arrival of each data point. 
These frequent updates could interfere with the timely provision of predictions 
for the incoming data.
Also, continuous learning is both computationally and economically expensive in terms of the computing infrastructure needed for the frequent re-learning of the data model.  

Here is why alternative learning strategies have been sought and the sliding window approach \cite{BravermanSlidingWindow} has caught on to deal with data streams.
In this learning modality, an ideal window $w$ of size $n$ slides over the incoming data stream to identify  batches (subsets) of data and the re-learning of the data model happens not at each data point but in accord with the sliding of $w$. This reduces both the number of data points used to update the data model
and the number of times the learning process is activated. 
This usually results in an important reduction in the overall computational cost.

Naturally a compromise has to be accepted between the model's  accuracy rate and 
its computational cost when substituting sliding window learning for continuous learning. 
The first shows, in general, a better computational cost whereas the seconds shows
a better accuracy.
In fact, when concept drift detectors (c.d.d.) are used together with a sliding window approach,  
the periodic re-learning of the model happens only when a concept drift is detected. 
Thus causing a late (from the ideal time point) re-training of the data model
which may result in a lower accuracy. 
Please recall that a concept drift is usually detected after the learning system realizes
the occurrence either of a {\em stable} increase in the model's error rate or 
of a {\em stable} variation in the statistical properties of the data stream.

Current research efforts in the area of concept drift detection can be grouped in:
\begin{itemize}
\item 
developing general purpose concept drift detectors 
\cite{conceptdriftandalgo, eddm, adwin, ddm,kswin,ph,hddm}.
Here the focus is in proving the general applicability of 
new c.d.d. usually using artificially generated data streams,
and to assess their complexity bounds.
Works belonging to this group tends to consider each c.d.d. as a stand-alone algorithm.
\item
performance assessment of learning systems exploiting c.d.d. evaluated as a whole  \cite{cavalcanteConceptDrift,gamaStreamLA}. And our present paper.
\item 
integrating c.d.d with a particular type of learners. For instance \cite{hultenDT, blackTsar, HoeglingerConceptDrift} integrate c.d.d. with decision tree learning. 
\item
ensemble approaches to concept drift detection. Here sets of data models and weighting schemes, or other selection mechanisms, are used to adapt the prediction function in the course of time \cite{wangConceptDriftEnsemble,streetConceptDrift}.
\end{itemize}

In this study, we propose a novel approach to do research on concept drift detection: we believe that c.d.d. need to be tailored, studied, and evaluated in relation to a specific domain. For our previous interests, we chose financial time series as investigative domain.

\section{Novel and standard concept drift detectors \label{scdd}}
Three concept drift detectors, specifically tailored to financial time series, have been developed and assessed in the paper. They are:
\begin{enumerate}
\item myTanDD - the detector calculates the angle between the tangent to the data and the x axis which is assumed at zero degree. A +80 degree angle represents an almost vertical line passing from the (0,0) point with increasing data values. A -80 degree angle represents an almost vertical line, passing from the (0,0) point, with decreasing data values. myTanDD uses the following threshold values to make a decision about the occurrence of a concept drift:
\begin{enumerate}
\item between -6 and +6 degrees, the time series is classified 'flat'\footnote{The threshold was fixed to 6 degrees following our definition of a bull market when using this concept drift detector. In particular a 5.71 degree slope corresponds to a 10\% earning rate as measured over a time series. We choose to define a bull market as a one where the slope of the time series is rising more than 10\% over a period of time thus the 6 degree threshold selection. The opposite definition holds for the bear market.
Finally, one can recall that it is common practice for learning processes to have some parameters set by preliminary and exploratory experimentation.}
\item higher than +6 degrees, the time series is classified 'bull' 
\item lower than -6 degrees, the time series is classified 'bear'
\item any time the time series changes classification a concept drift is detected.
\end{enumerate}
\item MINPS - the detector keeps track of two values: a) the data mean ($p$), calculated over all the data available at the current time, and b) the minimum standard deviation ($s_{min}$) from the set of all the standard deviations calculated at each time point up to the current time. 
Thus whereas $p$ may increase or decrease over time, $s_{min}$ can only stay constant or decrease in the course of time. A concept drift is detected when a data point falls outside the $p\pm3s_{min}$ interval. 
\item mySD - the detector calculates the standard deviation for all available data at each time point. The minimum value is stored in $s_{min}$. A concept drift is detected when the current standard deviation raises above $3 * s_{min}$.
\end{enumerate}
\noindent
It should be noted that these three detectors need to collect few data points (data window $dd\_window$) before starting to detect a concept drift to initialize their parameters and to provide statistical support to their decision. Also, the data window $dd\_window$ plus the current input represent all the data available to the three detectors in order to calculate their internal statistics and make a decision about the occurrence of a concept drift. When $dd\_window$ is full the oldest data point is removed and the current input appended.  
In all experiments reported, the data window's length is constant at 20 data points.  

Please also note that the term 'data' used in the three detectors' descriptions is a generic term and  refers to any input passed to the concept drift detector. In the following experiments, 'data' can be instantiated either as the model's accuracy or as a value taken from the incoming data stream (the 'raw' financial time series).

Finally, the standard concept drift detectors that we will consider in the empirical study are: EDDM \cite{eddm}, ADWIN \cite{adwin}, DDM \cite{ddm}, KSWIN \cite{kswin}, PH \cite{ph}, HDDM\_A \cite{hddm}, HDDM\_W \cite{hddm}. We please refer the interested reader to their reference papers for their detailed description.

\section{Novel and standard learning regressors \label{sls} }
As learning systems we considered some standard regressors: Linear Regression \cite{linregr}, Bayesian Ridge Regressor \cite{brr}, and MLPR \cite{mlpr}. They have been selected to provide a varied learning context where concept drift detectors could be evaluated.
In addition we exploited two basic regressors specifically designed for financial time series. They are:
\begin{enumerate}
\item YC (Yesterday Close) which returns the data stream value at time (t-1) as prediction for the value at time (t)
\item MinValInTS (Minimum Value In Training Set) which stores the minimum value in the training set 
and returns it for any future prediction.
\end{enumerate}
The definitions of the YC and MinValInTS regressors may appear trivial but in fact they are not. The YC regressor is based on the well-known observation that percentage variations among consecutive points in financial time series usually tends to be low and that the best prediction for the next day close is today close in absence of any specific information. 

The MinValTS regressor, instead, has been created in order to introduce an artificial concept drift (divergence) among the original financial time series and that made up by the predicted values. Thus allowing to monitor the behavior of the investigated concept drift detectors in presence of a permanent drift in the prediction time series.

\section{Learning with a sliding window: just in time concept drift detection \label{sliding}}
\begin{figure}
	\centering
	\includegraphics[width=1\linewidth]{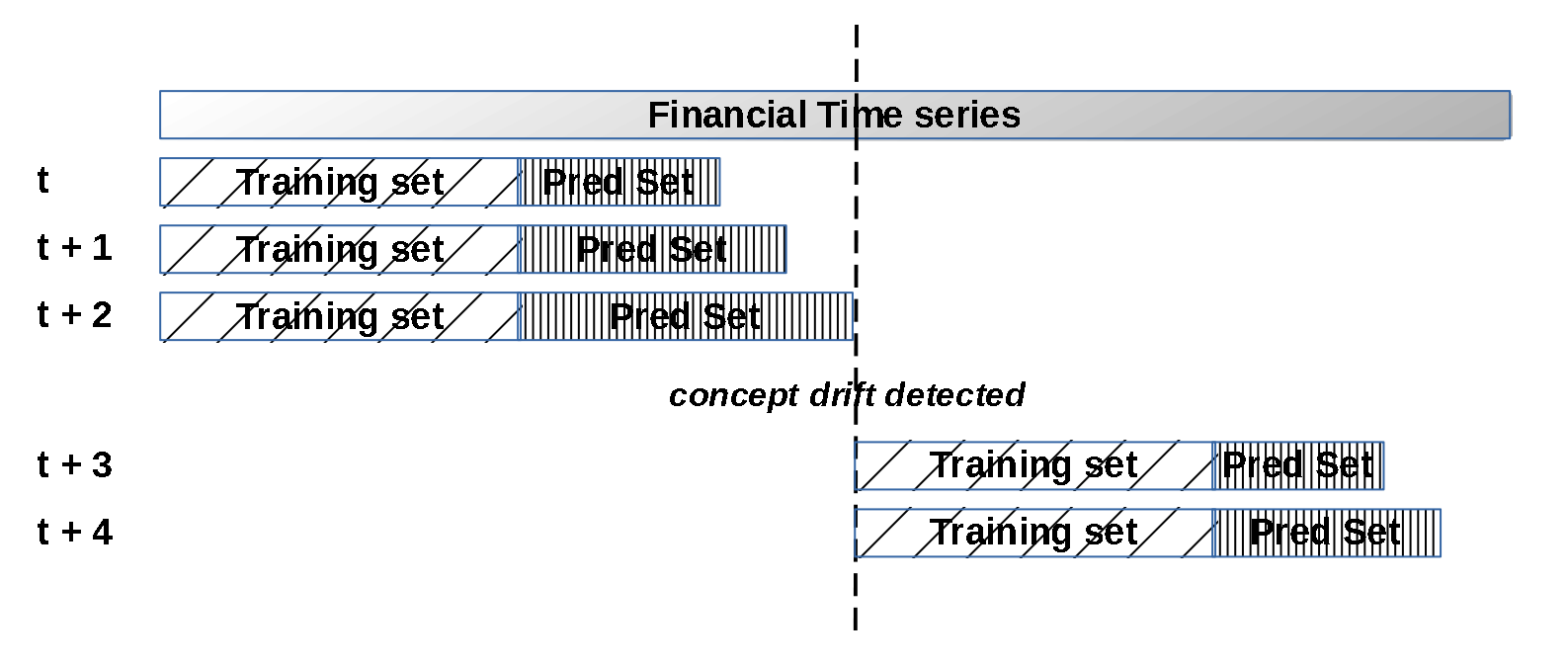}
	\caption{Learning with sliding window and concept drift detection. 
		The re-training step, after a concept drift is detected, is shown.}
	\label{cddtimeline}
\end{figure}

In this study, a sliding window learning approach together with a concept drift detector 
is employed, as it is usually done when a learning system is applied to a potential infinite, non-stationary data stream. 
Fig. \ref{cddtimeline} shows how the sliding window is implemented in our study. 
A number of data points is collected until a sufficiently long training set is formed (in our experiments, at least 30 data points). The learning system is run and a predictive model $M_1$ is built. 
Thereafter, $M_1$ is used to predict the next incoming data point. The last prediction is added to the prediction set which  grows indefinitely until a concept drift is detected.
Finally, either the incoming data point or the current model's accuracy is sent to the concept drift detector. 

In order to measure the data model's accuracy, we decided to use the Minimum Average Percentage Error (MAPE) \cite{maperef}. This is a common choice when one needs to quantify how much different two time series are \cite{NeriExSyst2020,Neri2019Wivace}.

In our experiments, we used two distinct MAPE functions. The first\\ $MAPE_{apd}(t)$ is calculated over all the predicted data points up to the time $t$ (the 'apd' term in the subscript means 'all predicted data').
This function is used to assess the overall performance of a learning system.
In fact, the $MAPE_{apd}(t)$ value calculated on the last data point in the time series is the one that we report in the experimental section.

The second $MAPE_{last\ k}(t)$ function (the term 'last k' in the subscript means 'last k data points in the prediction set) is calculated on the latest $k$ data points in the prediction set at time $t$ (or on the whole prediction set if shorter). 
In our experiments, k is fixed at 60. 
Then $MAPE_{last\ k}(t)$ measures the 'recent' accuracy of the model.
The values of $MAPE_{last\ k}(t)$ are then given to the concept drift detector in order to assess the occurrence of a concept drift.

The choice to use $MAPE_{last\ k}(t)$ as input to c.d.d. aims to avoid that recent changes in the time series (when a concept drift is emerging) are obfuscated by averaging over all the previous data points which likely include those generated when other concepts governed the time series. 

When a concept drift is detected, a new learning step begins: a new training set is collected and a novel model $M_2$ is learned. 
Fig. \ref{cddtimeline} shows that the novel training set is collected starting from the concept drift point. However, if the selected concept drift detector has signalled with a {\em warning point} the possible beginning of a concept drift before the concept drift point  was definitely detected, then the new training set will begin at the {\em warning point}.

Finally, we note that the performances of sliding window learning have to be compared with those of continuous learning in order to assess any significant differences.
We recall that when continuous learning is selected, the learning system begins a new learning step after predicting only one incoming data point. 
The comparison between the two learning modalities is discussed in the experimental section. 

\section{Average prediction error in financial time series \label{efs}}
\emph{Working Hypothesis: bounds on the average prediction error in financial time series. }
The best average prediction error at time $t$, $APE_{ls}(t)$, 
achievable by a learning system $ls$ applied to
a generic financial time series, 
in absence of other information, is included between the bounds reported in equation (\ref{errbound}): 
\begin{align}
    d_t \text{ is the data point at time t from ts} \\ 
    Diff_{ts}(t) = d_t - d_{t-1} \\ 
    AbsPerc_{ts}(t) = | \frac{Diff_{ts}(t)} {d_{t-1}} | \\
    avgAbsPercError_{ts}(t) = \frac{\sum_{1}^{t} AbsPerc_{ts}(t)} {t} 
\end{align}
\begin{equation}
    CorrectSign_{ts}(t) =
     \begin{cases}
      +1 & \text{if } d_t-d_{t-1} \geq 0 \\
      -1 & \text{if } d_t-d_{t-1} < 0 
     \end{cases}
\end{equation}
\begin{equation}
\begin{split}
    \hat{d}^{correct}_t  = &d_{t-1}* (1+ CorrectSign_{ts}(t)* \\
    & avgAbsPercError_{ts}(t)) \\
    \hat{d}^{wrong}_t = &d_{t-1}* (1+ -1*CorrectSign_{ts}(t)* \\
    & avgAbsPercError_{ts}(t))
\end{split}
\end{equation}
\begin{equation}
	\frac{\sum_{1}^{t} | \frac{\hat{d}^{correct}_{t} - d_{t-1}} {d_{t-1}} |} {t}   \leq 
	APE_{ls}(t) \leq 
    \frac{\sum_{1}^{t} | \frac{\hat{d}^{wrong}_{t} - d_{t-1}} {d_{t-1}} |} {t} \label{errbound}
\end{equation}	
\emph{Comment.}
To support our working hypothesis, this is not a proof, one can observe 
that at time $t$ some basic information available is: 
the past values of the time series $ts$ and how much each data point 
differs from the previous one in absolute percentage $AbsPerc_{ts}(t)$.

Therefore, at time $t$ a reasonable best guess about the next data point $d_{t+1}$ is to assume that 
it will be like $d_t$ apart an absolute average displacement equal to $AbsPerc_{ts}(t)$. 
And because no one knows if the next data value will be higher or lower than the current data point, 
then extreme estimates for $d_{t+1}$ could be $d_{t}*(1+AbsPerc_{ts})(t)$ and $d_{t}*(1-AbsPerc_)(t)$.
A reasonable interval of prediction for $d_{t+1}$ is then
\begin{equation}
 [ d_{t}*(1-AbsPerc_{ts}(t)), d_{t}*(1+AbsPerc_{ts}(t)) ].
 \end{equation}

Finally, if the learning system $ls$ is always right up to time $t$ about the direction of change between $d_{j-1}$ and $d_{j}$ for all $j \in (2...t)$,
its error rate will be $\frac{\sum_{1}^{t} | \frac{\hat{d}^{correct}_{t} - d_{t-1}} {d_{t-1}} |} {t}$.\\
Whereas if the learning system is always wrong up to time $t$ about the direction of change between $d_{j-1}$ and $d_{j}$ for all $j \in (2...t)$,
its error rate will be 
$\frac{\sum_{1}^{t} | \frac{\hat{d}^{wrong}_{t} - d_{t-1}} {d_{t-1}} |} {t}$.\\

This working hypothesis could be the explanation of why the YC (Yesterday Close) learner 
displays one of the lowest prediction error in our study.

To be noted that this working hypothesis does not prevent a learner to achieve a prediction rate lower
than the lower bound in equation (\ref{errbound}) by exploiting additional information
other that the single data points in the time series.

We are not aware of any research work that demonstrates this hypothesis, that is why we provide no reference, and we consider it a working hypothesis that in our past works on financial time series has always being verified \cite{NeriAIcom12,NeriExSyst2020,Neri2019Wivace}.

\section{Experimental setting \label{expSec}}
In order to perform the experiments, we implemented the sliding window learning process of Section \ref{sliding} in Python 3 and used the classic drift detectors listed in Section \ref{scdd} as implemented in the scikit-multiflow package \cite{skmultiflow}. The hardware used was a Dell XPS 13 notebook with an Intel i-7 CPU and 16 GB of RAM. 

The selected financial time series for the study were: 
\begin{itemize}
\item SPDR SP500 index (SPY), from Dec 14 2015 to Dec 11 20202, and
\item Bank ETF (KBE), from Jan 03, 2016 to Jan 03, 2021.
\end{itemize}
Their data are freely available on the Internet, for instance from finance.yahoo.com. 

In order to provide a minimal set of information so that the learning systems 
could try to learn a model of the time series, 
each instance given to the learners contains the following information: 
\emph{
\begin{tabbing}
AAA\= BBB \= \kill	
 \>instance at time t:  \\
 \>\>$<$close value(t-3), close value(t-2), \\
 \>\>\ close value(t-1), close value(t)$>$  
\end{tabbing}
}  
\noindent
where $close\ value(t)$ is the prediction target. 

During the experimental analysis, we examined every combination (configuration) of the selected concept drift detectors (described in Section \ref{scdd}) with the chosen learning systems (described in Section \ref{sls}) and the selected financial time series. 
And we collected a set of performance measures to understand how performant each configuration is. 
The continuous learning modality (Section \ref{sls}) is included in the configuration set. Note that when continuous learning is employed no concept drift detector is exploited as it would be meaningless.

Each experimental configuration has been labeled following the schema: 
{\em 
\begin{tabbing}
AAA\= BBB \= \kill	
	\>$<$learner$>$ $<$concept drift detector$>$\\
	\>$<$data given to the concept drift detector$>$ \\
	\>$<$if continuous learning is used $>$. 
\end{tabbing}
}
\noindent
Therefore, just to provide an example, in the following figures and tables, the experiment named 
'MinValInTS ADWIN MAPE contLearn F' represents the configuration with: learning system equal to MinValInTS, concept drift detector equal to ADWIN, ADWIN receives as input the current MinValInTS' error rate, calculated by using MAPE, and no continuous learning is used.

\begin{figure*}[tbh]
	\centering
	\includegraphics[width=1.02\linewidth]{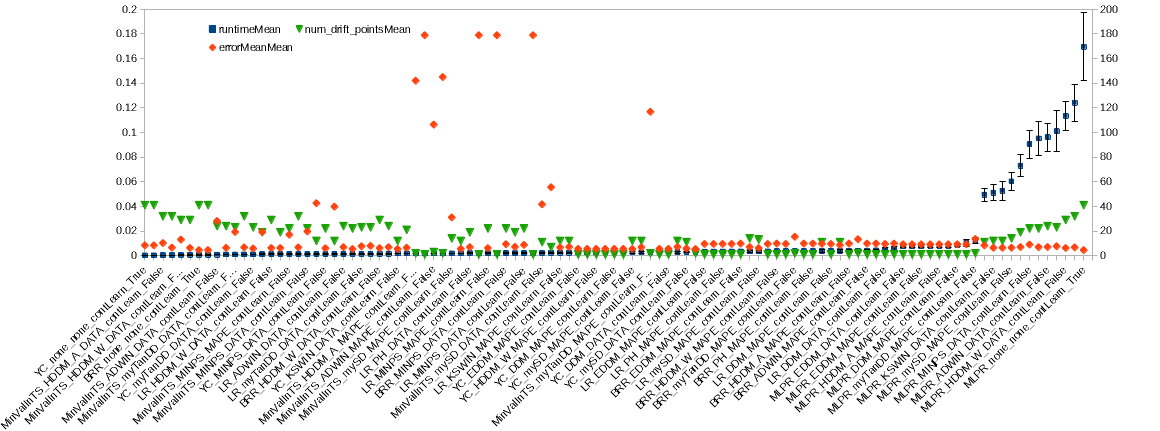}
	\caption[]{Time series SPY. Runtime in seconds (right axis), number of drifts detected (right axis), and MAPE (left axis) for all configurations ordered by runtime.}
	\label{spyrde}
\end{figure*}

\begin{figure*}[tbh]
	\centering
	\includegraphics[width=1.0\linewidth]{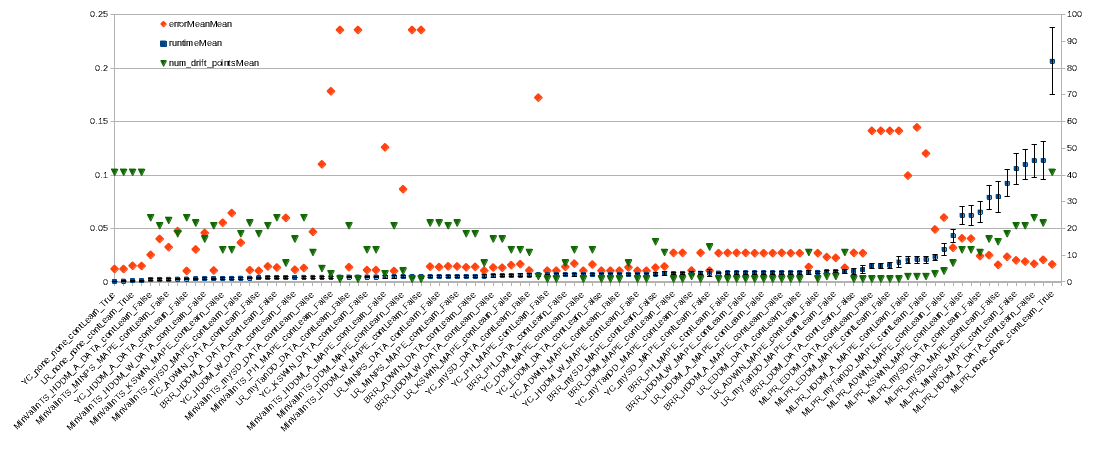}
	\caption[]{Time series KBE. Runtime in seconds (right axis), number of drifts detected (right axis), and MAPE (left axis) for all configurations ordered by runtime.}
	\label{kberde}
\end{figure*}

All the data reported in 
fig. \ref{spyrde} and \ref{kberde} and in Table \ref{tabSPY} and \ref{tabKBE} are averaged over 10 runs. The configurations (x axis) reported in fig. \ref{spyrde} and \ref{kberde} are ordered by their increasing runtime.

In Table \ref{tabSPY} and \ref{tabKBE}, the runtime, the error rate (MAPE), and the number of concept drifts detected is reported for all configurations and for the two selected financial time series. The configurations in the tables are ordered by their increasing MAPE value.

Fig. \ref{spyrde} and \ref{kberde} show that most configurations need a very short runtime for both financial time series. This is due to the short runtime of most learners. The only exception happens when using MLPR (Multi Layer Perceptron Regressor) which requires a relatively long learning time.
 
When a long runtime is needed for the learning step, the runtimes of the prediction and of the concept drift detection steps become marginal if measured as percentages over the total runtime.
In fact fig. \ref{mlprcomplexity} shows that the learning cost represents about 90\% of the runtime in the case of configurations with the highest runtimes.
This can be easily verified by matching the configurations in the right half of the barchart in fig. \ref{mlprcomplexity} with the configurations in the right end of fig. \ref{spyrde}. 
A similar distribution of runtimes and relative costs also appears when considering the KBE time series.

By comparing the error rate curve (MAPE) in fig. \ref{spyrde} and fig. \ref{kberde}, one can 
note that MAPE tends to be quite low across most configurations. And by taking advantage of the numeric data in Tables \ref{tabSPY} and \ref{tabKBE}, it can be observed that the configurations using the YC (Yesterday Close) learner tend to produce the lowest error rates whereas those using the MinValInTS learner produce the highest error rates. 
The first observation is coherent with the working hypothesis on the bounds of the average prediction 
error in financial time series that we discussed in Section \ref{efs}. The second observation is to be expected because the MinValInTS learner always predicts the same constant value thus introducing an ever-increasing prediction error unless the financial time series stays perfectly constant. 

The third curve in fig. \ref{spyrde} and \ref{kberde} shows the number of concept drifts detected. Tables \ref{tabSPY} and \ref{tabKBE} also report the information in a numeric format.

A surprising and unexpected result is the high variation in concept drift detections. Some c.d.d. just detect a single concept in the financial time series. Others detect the occurrence of several concepts. 

Reading the phenomena from the perspective of financial technical analysis, some concept drift detectors are sensitive to small variations occurring in the time series in short periods of time while others are only sensible to large variations over long periods of time. 

There is no good or bad when detecting concept drifts in financial time series in a real time setting. Even human experts are not able to identify significant changes of market phases in real time. From this point of view, the experiments are informative because they show the different sensitivity of the many concept drift detectors.

The experiments are particularly insightful from a practical point of view as they show which concept drift detector has contributed or not in keeping  the MAPE values as low  as possible in a given configuration. 
From Table \ref{tabSPY} and \ref{tabKBE}, one can observe that lowest MAPE values are correlated with a relatively high number of drift identifications. Thus a c.d.d. is good if it can find the time point at which the current model become outdated and needs to be re-trained.

In particular, the DDM and EDDM drift detectors, when applied to the data points in the financial time series, detect only one concept. Please refer to the configuration 'any\_learner DDM Data F' and 'any\_learner EDDM Data F' in the tables. 
This means that these two configurations are useless for detecting market phases in dealing with financial time series.
Unfortunately, none can be said for the configurations exploiting MAPE as time series given to the c.d.d. because the shape of the MAPE curve is a priori unknown and emerges dynamically only  when the learner starts to make predictions on the given time series.

\subsection{Finding Equivalent Configurations}
The configurations in the first half of Tables \ref{tabSPY} and \ref{tabKBE} show MAPE differences that are statistically significant. This would allow to rank some configurations as better than the others but only in the context of the performed experiments not for any time series. 
In fact the No Free Lunch Theorem (NFL Theorem) \cite{nofreelunch} reminds that no learning systems can outperform all the others under any operating conditions.

However, we believe that the methodology used to construct Tables \ref{tabSPY} and \ref{tabKBE} has general validity and can be used to identify a set of almost equivalent configurations, in terms of performances,
for a given financial time series (thus respecting the NFL Theorem).

The methodology {\em Find\_Equivalent\_Configuration\_Set} works as follows:
\begin{enumerate}
\item given a time series $ts$ and a set of configurations $C_{init}$ =
 \{$<$learner, concept drift detector, data given to the c.d.d.$>$\} 
\item include among the learners our YC learner and update $C_{init}$ accordingly  
\item build a data table like Table \ref{tabSPY} 
\item assume that the working hypothesis in Section \ref{efs} holds 
\item $ref\_error$ =  $min( {MAPE(config) | config \in C_{init} } )$
\item assume equivalent all the configurations $C_{init}$ whose MAPE value 
is $\leq k * ref\_error$, where $k>1$ is chosen by the human experimenter, 
and include them in $C_{equiv}$
\end{enumerate}
 
Following the {\em Find\_Equivalent\_Configuration\_Set} methodology in our study, 
one could select MAPE('BRR none none T') for the SPY time series and
MAPE('BRR none none T') for the KBE time series as reference errors ($ref\_error$).
$C_{equiv}$ would then include those 
configurations whose MAPE value is $\leq k * MAPE(ref\_error)$.
In the case of our experiments, it seems reasonable to select $k=2$ considering how small are 
the MAPE values of the reference configurations. Thus $C_{equiv}$ for the two time series would contain all configurations in the first part of Table \ref{tabSPY} or \ref{tabKBE} respectively. 

\subsection{Selecting concept drift detection configuration in practice}
We show here how to select the best performing concept drift detectors configurations for a given time series.
By applying the methodology {\em Find\_Equivalent\_Configuration\_Set} above, one can calculate 
the set of equivalent configurations $C_{equiv}$ for the time series.
If L is the set of all learners $l_i$ in $C_{equiv}$,
then the best concept drift configuration set $C_{best}(ts)$ can be defined as:
\begin{equation}
\begin{split}
C_{best}(ts) = &\{ <any, cdd, input, F> | \\
  &\ for \ every \ l \  \in \ L,\\
  & \ <l, cdd, input, F>\  \in \ C_{equiv} \} 
\end{split}
\end{equation}
In words, $C_{best}(ts)$ contains the 4-tuple 
\begin{center}
	 $<$ any, cdd, input, F $>$ 
\end{center} 
only if 
the pair $<$ cdd, input $>$ appears among the configurations of each learners in $C_{equiv}$. 

According to these definitions and the data in Tables \ref{tabSPY} and \ref{tabKBE}, 
it is possible to calculate the sets
$C_{best}(SPY)$ and $C_{best}(KBE)$. They are listed without any specific order 
in Table \ref{bestCDD}.
\begin{table}[htb]
\centering
\begin{tabular}{|c | c|}
 \hline
 {\bf $C_{best}(SPY)$} 			& {\bf $C_{best}(KBE)$ } \\
 \hline
 any, HDDM\_A, DATA, F		& any, HDDM\_A, DATA, F \\
 any, PH, DATA, F				& any, PH, DATA, F \\
 any, ADWIN, MAPE, F			& any, ADWIN, MAPE, F \\
 any, MINPS, MAPE, F			& any, MINPS, MAPE, F \\
 any, mySD, MAPE, F			& any, mySD, MAPE, F \\
 any, KSWIN, DATA, F			& any, MINPS, DATA, F \\
 any, KSWIN, MAPE, F			& any, mySD, DATA, F \\
							& any, HDDM\_W, DATA, F \\
 \hline
\end{tabular}
\caption{Time series SPY and KBE. Their best performing concept drift detection configurations. No order exists among the listed configurations.} 
\label{bestCDD}
\end{table} 

The configurations containing the pairs in $C_{best}(SPY)$ and $C_{best}(KBE)$ display
runtimes that are significantly lower than the runtimes of their relative  
continuous learning configurations.
For instance, on the SPY time series the configuration 'MLPR none none cL T' requires 169.47 seconds whereas
'MLPR mySD MAPE cL F only 60.33 seconds and 'MLPR HDDM\_A DATA cL F' 124.14 seconds.
Naturally the saving in runtime is more evident when the learner's execution takes a long time.

By applying the above methods to future time series $ts$, one could easily identify both the set of equivalent configurations $C_{equiv}$ and the set of the best (concept drift detection) configurations $C_{best}(ts)$ that exhibit a stable and consistent behavior over the time series for all learners.
This is a practical and operative result.
 
\begin{figure}[tb]
	\centering
	\includegraphics[width=1\linewidth]{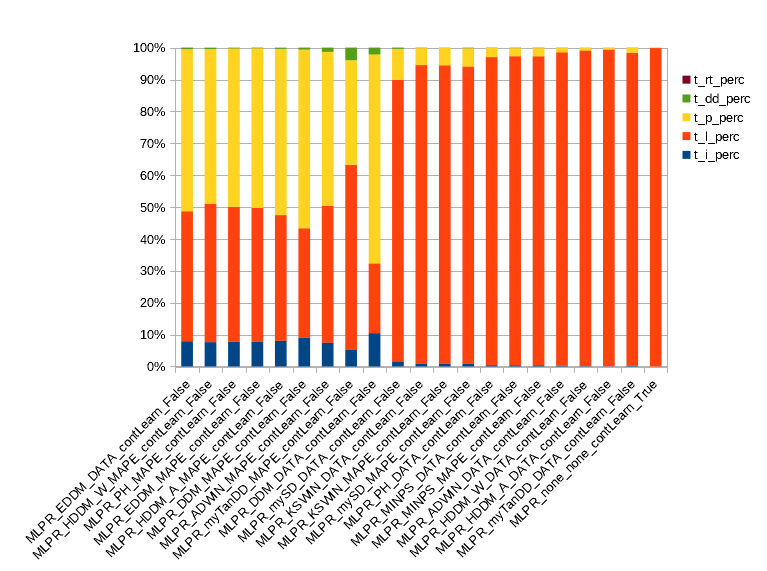}
	\caption{Time series SPY. Computational costs, in percentage, for the main components of our configurations: learning (t\_l\_perc), prediction (t\_p\_perc), drift detection (t\_dd\_perc), updating the service data structures (t\_i\_perc and t\_rt\_perc). The barchart reports all the configurations with MLPR as a learner ordered by their total runtime.}
	\label{mlprcomplexity}
\end{figure}

\begin{table}[htb]
	\fontsize{6}{5}\selectfont
	\begin{filecontents*}{data/SPYedr.csv}
config,error,eStd,runtime,rStd,ndp
BRR_none_none_cL_T,0.0046,0,0.83,0.29,41
LR_none_none_cL_T,0.0046,0,0.84,0.38,41
MLPR_none_none_cL_T,0.0046,0,169.47,27.6,41
YC_KSWIN_DATA_cL_F,0.0055,0.0001,1.87,0.23,12
YC_MINPS_MAPE_cL_F,0.0055,0,1.58,0.37,22
YC_mySD_MAPE_cL_F,0.0055,0,2.74,0.47,12
YC_myTanDD_MAPE_cL_F,0.0055,0,3.16,0.34,2
YC_EDDM_MAPE_cL_F,0.0056,0,2.59,0.28,1
YC_PH_MAPE_cL_F,0.0056,0,2.63,0.35,1
YC_HDDM_W_MAPE_cL_F,0.0056,0,2.64,0.26,1
YC_EDDM_DATA_cL_F,0.0056,0,2.65,0.26,1
YC_DDM_MAPE_cL_F,0.0056,0,2.66,0.31,1
YC_HDDM_A_MAPE_cL_F,0.0056,0,2.7,0.3,1
YC_ADWIN_MAPE_cL_F,0.0056,0,2.8,0.29,1
YC_DDM_DATA_cL_F,0.0056,0,2.84,0.33,1
YC_KSWIN_MAPE_cL_F,0.0056,0.0002,1.95,0.27,12
YC_mySD_DATA_cL_F,0.0058,0,3.03,0.56,11
YC_myTanDD_DATA_cL_F,0.0059,0,1.08,0.16,23
YC_MINPS_DATA_cL_F,0.0061,0,1.43,0.2,22
BRR_HDDM_W_DATA_cL_F,0.0062,0,1.65,0.45,29
LR_MINPS_MAPE_cL_F,0.0062,0,2.05,0.39,22
LR_mySD_MAPE_cL_F,0.0062,0,3.51,0.66,13
YC_HDDM_W_DATA_cL_F,0.0063,0,0.78,0.15,29
LR_HDDM_W_DATA_cL_F,0.0063,0,1.28,0.25,29
YC_PH_DATA_cL_F,0.0063,0,1.32,0.16,19
MLPR_HDDM_W_DATA_cL_F,0.0063,0,113.5,11.58,29
YC_ADWIN_DATA_cL_F,0.0064,0,0.97,0.14,24
MLPR_KSWIN_DATA_cL_F,0.0065,0.0005,51.17,6.3,12
MLPR_KSWIN_MAPE_cL_F,0.0065,0.0003,52.73,7.8,12
BRR_MINPS_MAPE_cL_F,0.0065,0,1.88,0.35,21
MLPR_mySD_MAPE_cL_F,0.0065,0,60.33,6.94,14
YC_HDDM_A_DATA_cL_F,0.0066,0,0.54,0.11,32
LR_KSWIN_MAPE_cL_F,0.0068,0.0008,2.49,0.44,12
LR_HDDM_A_DATA_cL_F,0.0068,0,1.08,0.35,32
BRR_HDDM_A_DATA_cL_F,0.0068,0,1.35,0.53,32
MLPR_HDDM_A_DATA_cL_F,0.0068,0,124.14,14.93,32
BRR_KSWIN_MAPE_cL_F,0.0069,0.0006,2.75,0.71,12
LR_ADWIN_DATA_cL_F,0.007,0,1.44,0.21,24
BRR_ADWIN_DATA_cL_F,0.007,0,1.7,0.41,24
LR_PH_DATA_cL_F,0.007,0,2,0.29,19
MLPR_MINPS_MAPE_cL_F,0.007,0,95.15,13.95,22
BRR_mySD_MAPE_cL_F,0.0071,0,3.51,1.12,14
MLPR_PH_DATA_cL_F,0.0071,0,73.08,9.07,19
MLPR_ADWIN_DATA_cL_F,0.0071,0,96.15,11.24,24
LR_KSWIN_DATA_cL_F,0.0071,0.0006,2.53,0.36,12
BRR_KSWIN_DATA_cL_F,0.0072,0.0007,2.86,1.03,12
BRR_PH_DATA_cL_F,0.0072,0,2.13,0.37,19
LR_myTanDD_DATA_cL_F,0.0077,0,1.63,0.26,23
MLPR_myTanDD_DATA_cL_F,0.0077,0,101.04,16.63,23
BRR_myTanDD_DATA_cL_F,0.008,0,1.64,0.42,23
YC_none_none_cL_T,0.0084,0,0.31,0.1,41
MinValInTS_none_none_cL_T,0.0084,0,0.34,0.09,41
MLPR_mySD_DATA_cL_F,0.0084,0,49.55,5.24,11
LR_mySD_DATA_cL_F,0.0086,0,3.94,1.02,11
LR_MINPS_DATA_cL_F,0.0088,0,2.16,0.48,22
MLPR_MINPS_DATA_cL_F,0.0091,0,90.51,11.55,22
MLPR_EDDM_DATA_cL_F,0.0093,0,7.61,0.85,1
MLPR_PH_MAPE_cL_F,0.0093,0,7.68,0.86,1
MLPR_EDDM_MAPE_cL_F,0.0093,0,7.74,0.89,1
MLPR_HDDM_W_MAPE_cL_F,0.0093,0,7.85,0.89,1
MLPR_HDDM_A_MAPE_cL_F,0.0093,0,7.95,0.95,1
MLPR_ADWIN_MAPE_cL_F,0.0093,0,8.01,1.13,1
MLPR_DDM_MAPE_cL_F,0.0093,0,8.68,1.35,1
MLPR_DDM_DATA_cL_F,0.0093,0,10.57,2.76,1
BRR_MINPS_DATA_cL_F,0.0094,0,2.11,0.4,22
LR_EDDM_MAPE_cL_F,0.0096,0,3.34,0.4,1
LR_HDDM_W_MAPE_cL_F,0.0096,0,3.39,0.66,1
LR_PH_MAPE_cL_F,0.0096,0,3.43,0.46,1
LR_HDDM_A_MAPE_cL_F,0.0096,0,3.45,0.65,1
LR_EDDM_DATA_cL_F,0.0096,0,3.56,0.5,1
LR_ADWIN_MAPE_cL_F,0.0096,0,3.72,0.62,1
LR_DDM_MAPE_cL_F,0.0096,0,3.93,0.74,1
LR_DDM_DATA_cL_F,0.0096,0,5.1,1.72,1
BRR_mySD_DATA_cL_F,0.0097,0,3.89,1.24,11
BRR_EDDM_MAPE_cL_F,0.0099,0,3.5,0.56,1
BRR_HDDM_W_MAPE_cL_F,0.0099,0,3.71,0.68,1
BRR_EDDM_DATA_cL_F,0.0099,0,3.79,0.84,1
BRR_PH_MAPE_cL_F,0.0099,0,3.8,0.85,1
BRR_HDDM_A_MAPE_cL_F,0.0099,0,4.03,0.77,1
BRR_ADWIN_MAPE_cL_F,0.0099,0,4.1,1.29,1
BRR_DDM_MAPE_cL_F,0.0099,0,4.35,0.94,1
BRR_DDM_DATA_cL_F,0.0099,0,5.59,2.13,1
MinValInTS_HDDM_A_DATA_cL_F,0.0103,0,0.47,0.14,32
MinValInTS_HDDM_W_DATA_cL_F,0.0131,0,0.67,0.19,29
LR_myTanDD_MAPE_cL_F,0.0134,0,4.05,1.18,2
MLPR_myTanDD_MAPE_cL_F,0.0136,0,12.43,2.26,2
BRR_myTanDD_MAPE_cL_F,0.0154,0,3.72,0.73,2
MinValInTS_MINPS_MAPE_cL_F,0.0171,0,1.34,0.23,22
MinValInTS_PH_DATA_cL_F,0.0192,0,1.2,0.33,19
MinValInTS_myTanDD_DATA_cL_F,0.0193,0,0.99,0.17,23
MinValInTS_MINPS_DATA_cL_F,0.0197,0,1.38,0.25,22
MinValInTS_ADWIN_DATA_cL_F,0.0282,0,0.86,0.08,24
MinValInTS_mySD_MAPE_cL_F,0.0311,0,1.95,0.21,14
MinValInTS_KSWIN_DATA_cL_F,0.0398,0.0014,1.43,0.08,12
MinValInTS_mySD_DATA_cL_F,0.0417,0,2.3,0.26,11
MinValInTS_KSWIN_MAPE_cL_F,0.0426,0.0003,1.39,0.1,12
MinValInTS_DDM_MAPE_cL_F,0.0556,0,2.45,0.47,7
MinValInTS_ADWIN_MAPE_cL_F,0.1064,0,1.93,0.15,3
MinValInTS_myTanDD_MAPE_cL_F,0.1168,0,2.76,0.44,2
MinValInTS_HDDM_A_MAPE_cL_F,0.142,0,1.88,0.18,2
MinValInTS_PH_MAPE_cL_F,0.145,0,1.94,0.23,2
MinValInTS_HDDM_W_MAPE_cL_F,0.1789,0,1.9,0.34,1
MinValInTS_EDDM_DATA_cL_F,0.1789,0,2.01,0.2,1
MinValInTS_EDDM_MAPE_cL_F,0.1789,0,2.07,0.25,1
MinValInTS_DDM_DATA_cL_F,0.1789,0,2.18,0.33,1
\end{filecontents*}
\csvautotabular[respect all]{data/SPYedr.csv}
	\par
	\caption{Time series SPY. Runtime with standard deviation, number of drifts detected with standard deviation, and MAPE for all configurations ordered by the error mean (MAPE). The number of drift point series shows zero standard deviation for all cases.} 
	\label{tabSPY}
\end{table}

\begin{figure}[tb]
	\centering
	\includegraphics[width=0.9\linewidth]{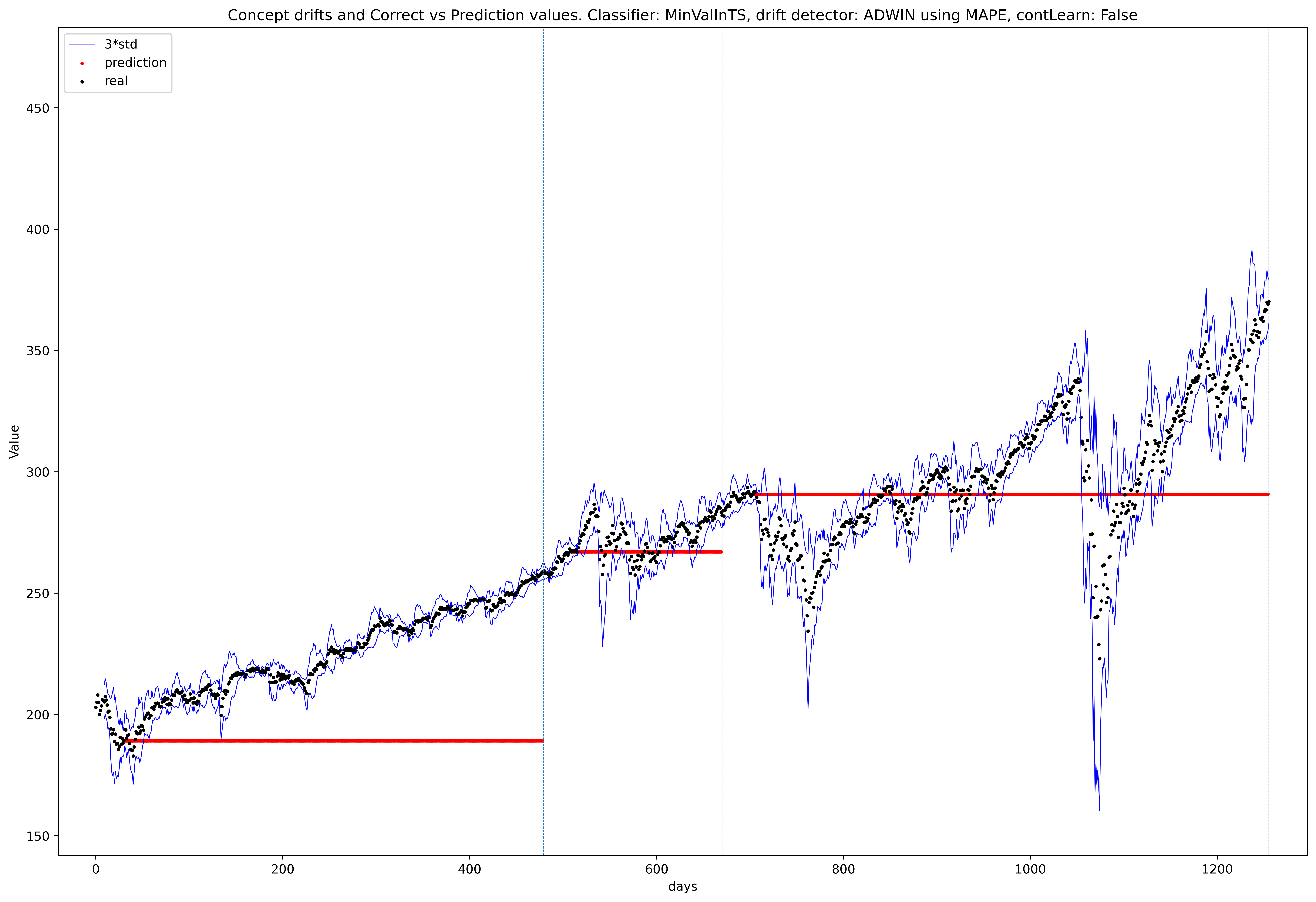}
	\caption[]{Time series SPY. Real and prediction curves for one single experiment in configuration: 'MinValInTS ADWIN MAPE cL F'.}
	\label{spypred}
\end{figure}

\begin{figure}[tb]
	\centering
	\includegraphics[width=0.9\linewidth]{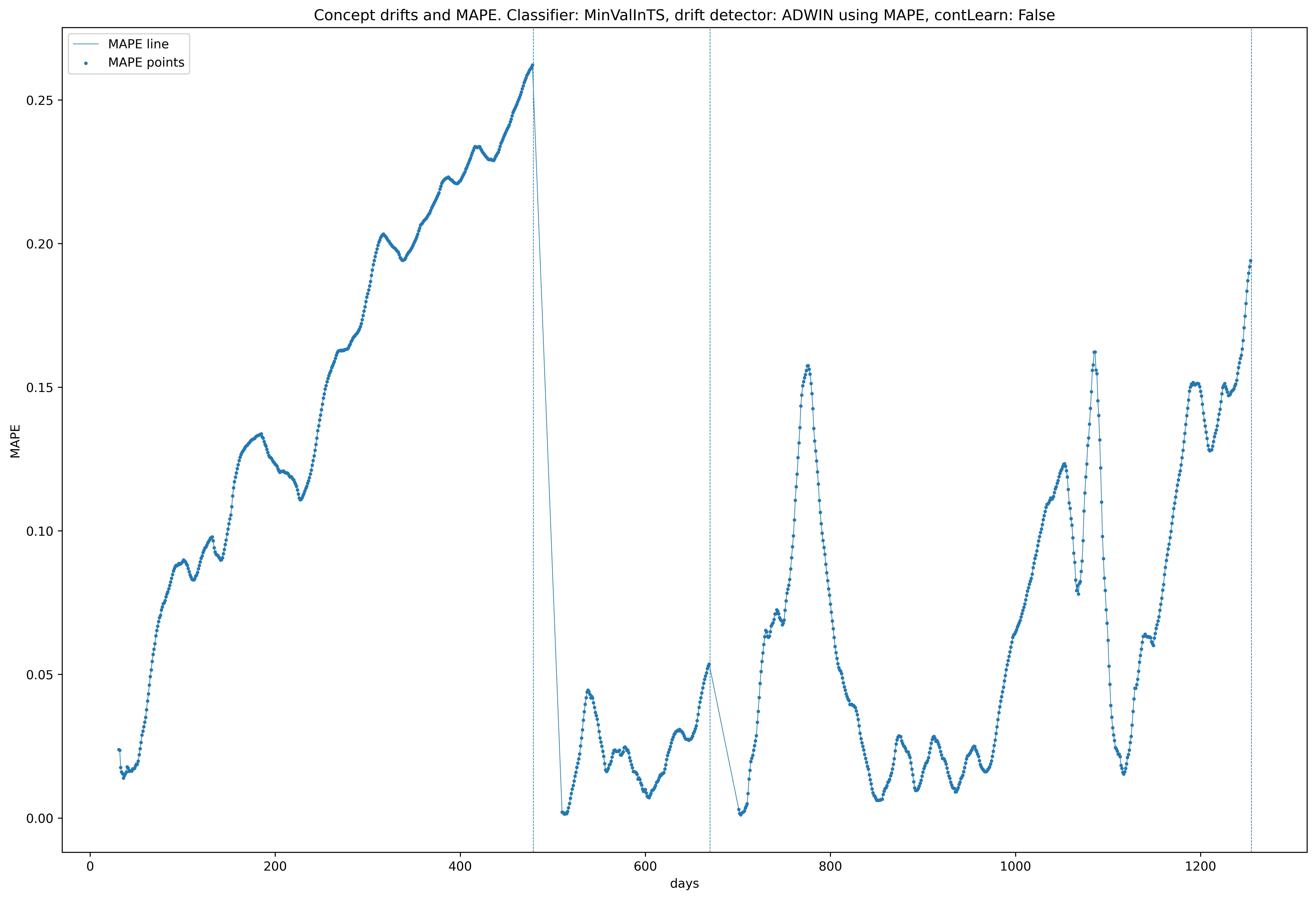}
	\caption[]{Time series SPY. MAPE curve for one single experiment in configuration: 'MinValInTS ADWIN MAPE cL F'. }
	\label{spyerror} 
\end{figure}

\begin{figure}[t]
	\centering
	\includegraphics[width=0.9\linewidth]{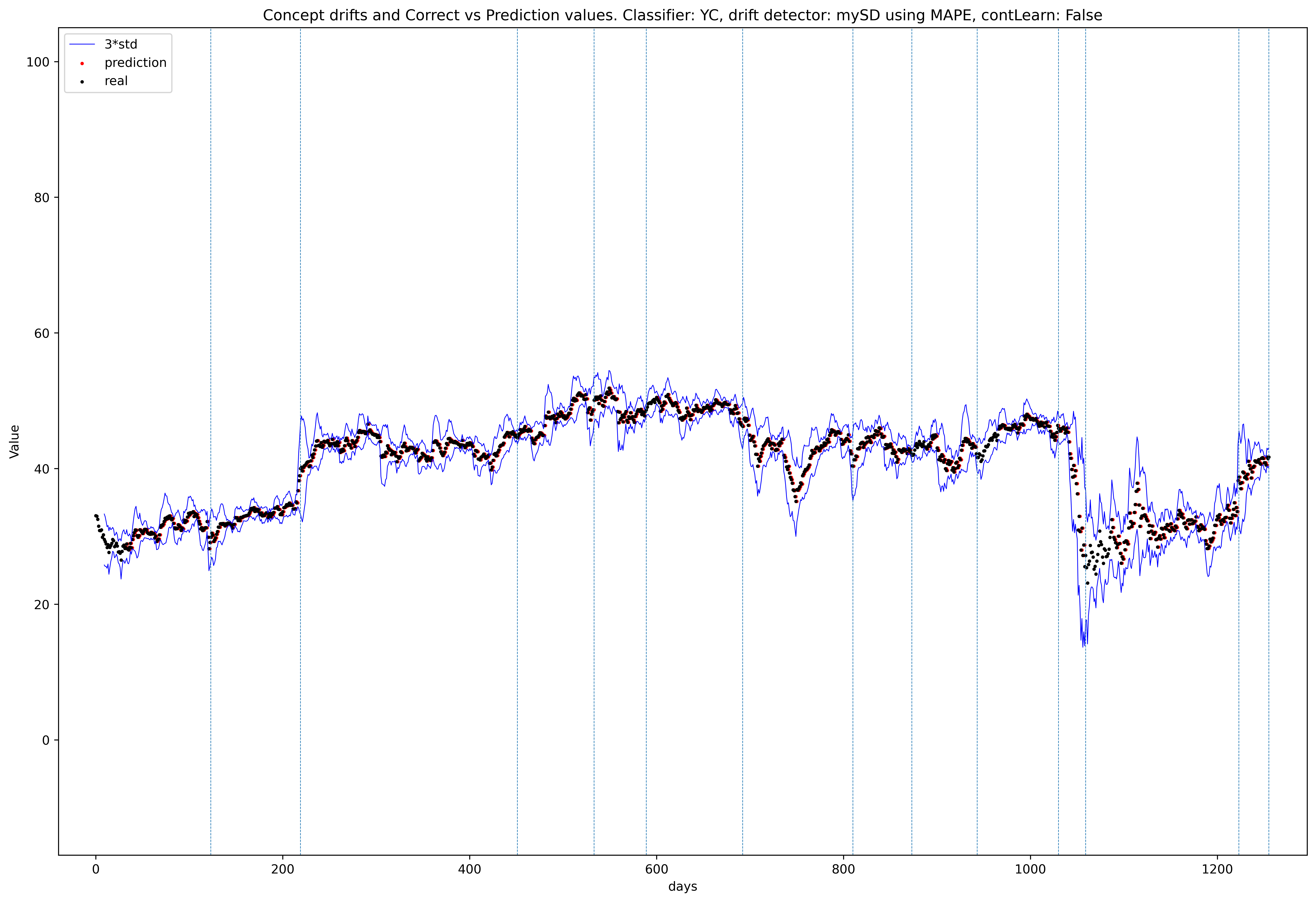}
	\caption[]{Time series KBE. Real and prediction curves for one single experiment in configuration: 'YC mySD MAPE cL F'.}
	\label{kbepred}
\end{figure}

\begin{figure}[tb]
	\centering
	\includegraphics[width=0.9\linewidth]{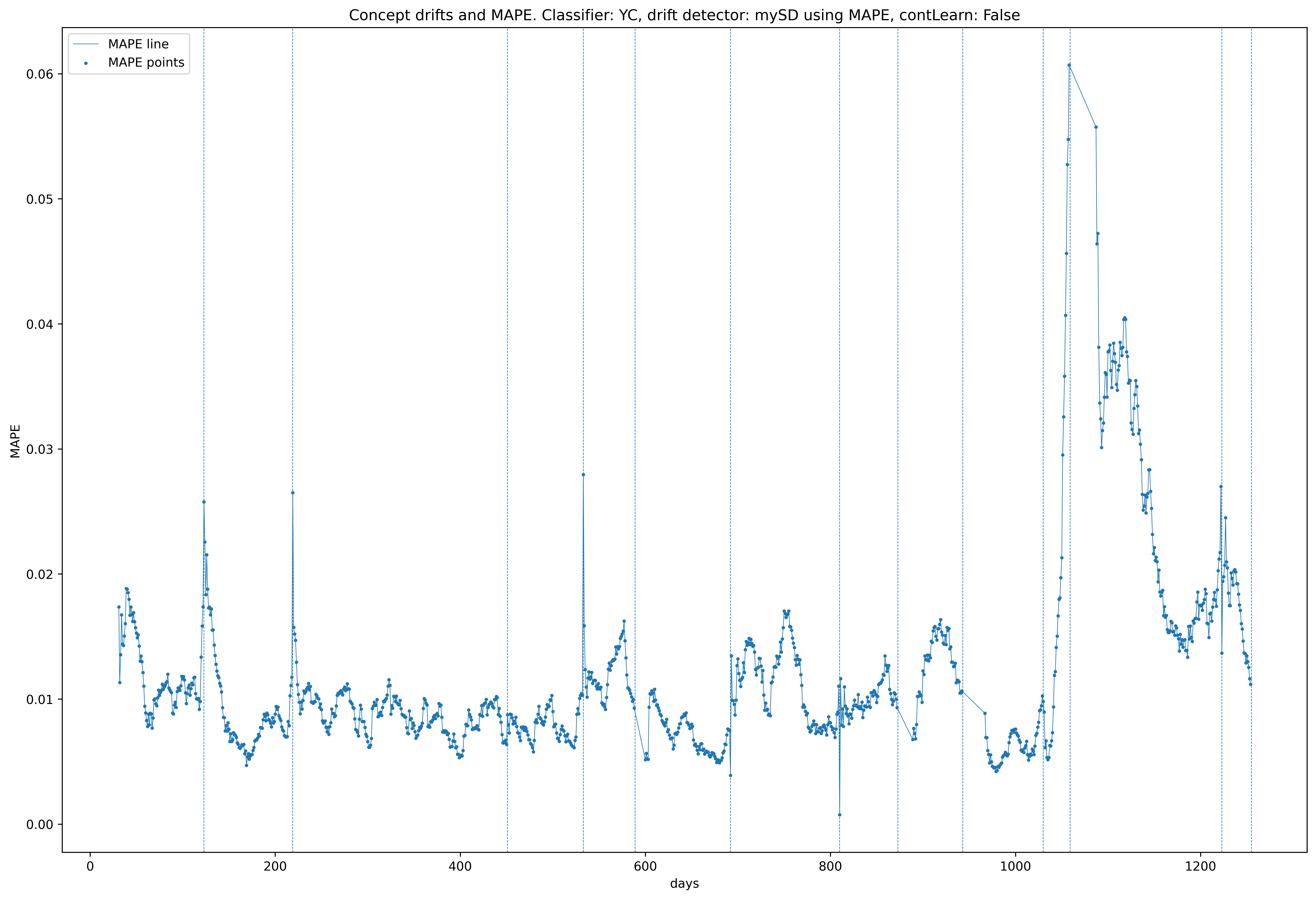}
	\caption[]{Time series KBE. MAPE curve for one single experiment in configuration: 'YC mySD MAPE cL F'.}
	\label{kbeerror}
\end{figure}

\begin{table}[htb]
	\fontsize{6}{5}\selectfont
	\begin{filecontents*}{data/KBEedr.csv}
config,error,eStd,runtime,rStd,ndp
YC_MINPS_MAPE_cl_F,0.0102,0,1.87,0.55,21
YC_HDDM_A_DATA_cl_F,0.0103,0,1.1,0.18,24
YC_ADWIN_DATA_cl_F,0.0104,0,1.51,0.3,18
YC_mySD_MAPE_cl_F,0.0105,0,3.33,1.25,13
YC_DDM_MAPE_cl_F,0.0105,0,2.72,0.41,1
YC_DDM_DATA_cl_F,0.0105,0,2.77,0.49,1
YC_EDDM_MAPE_cl_F,0.0105,0,2.7,0.4,1
YC_EDDM_DATA_cl_F,0.0105,0,2.75,0.47,1
YC_PH_MAPE_cl_F,0.0105,0,2.67,0.27,1
YC_ADWIN_MAPE_cl_F,0.0105,0,2.78,0.47,1
YC_HDDM_A_MAPE_cl_F,0.0105,0,2.87,0.34,1
YC_HDDM_W_MAPE_cl_F,0.0105,0,2.82,0.52,1
YC_myTanDD_MAPE_cl_F,0.0106,0,3.24,0.64,2
YC_PH_DATA_cl_F,0.0106,0,2.2,0.44,7
YC_myTanDD_DATA_cl_F,0.0107,0,1.27,0.22,21
YC_mySD_DATA_cl_F,0.0107,0,2.52,0.51,11
YC_MINPS_DATA_cl_F,0.011,0,1.45,0.24,22
YC_KSWIN_DATA_cl_F,0.011,0.0001,1.85,0.21,12
YC_KSWIN_MAPE_cl_F,0.011,0.0001,1.86,0.22,12
YC_HDDM_W_DATA_cl_F,0.0114,0,1.67,0.14,16
MinValInTS_none_none_cl_T,0.0121,0,0.33,0.13,41
YC_none_none_cl_T,0.0121,0,0.17,0.06,41
LR_HDDM_A_DATA_cl_F,0.013,0,1.72,0.46,24
LR_mySD_MAPE_cl_F,0.0132,0,3.91,0.88,11
LR_HDDM_W_DATA_cl_F,0.0133,0,2.4,0.53,16
BRR_mySD_MAPE_cl_F,0.0134,0,2.93,0.69,15
BRR_HDDM_A_DATA_cl_F,0.0135,0,1.64,0.39,24
BRR_HDDM_W_DATA_cl_F,0.0135,0,2.35,0.53,16
LR_PH_DATA_cl_F,0.0136,0,2.79,0.31,7
LR_ADWIN_DATA_cl_F,0.0137,0,2.04,0.31,18
LR_myTanDD_DATA_cl_F,0.0138,0,1.85,0.26,21
LR_MINPS_DATA_cl_F,0.0139,0,2.02,0.24,22
LR_mySD_DATA_cl_F,0.014,0,3.51,0.66,11
BRR_PH_DATA_cl_F,0.0141,0,2.71,0.38,7
BRR_ADWIN_DATA_cl_F,0.0141,0,2.07,0.44,18
BRR_MINPS_DATA_cl_F,0.0143,0,1.96,0.44,22
LR_MINPS_MAPE_cl_F,0.0145,0,2.02,0.43,22
BRR_myTanDD_DATA_cl_F,0.0145,0,1.63,0.26,21
BRR_mySD_DATA_cl_F,0.0145,0,3.09,0.68,11
BRR_MINPS_MAPE_cl_F,0.0146,0,2.02,0.38,21
BRR_none_none_cl_T,0.0149,0,0.52,0.16,41
LR_none_none_cl_T,0.0151,0,0.51,0.19,41
LR_KSWIN_MAPE_cl_F,0.0157,0.0006,2.43,0.34,12
MLPR_mySD_MAPE_cl_F,0.016,0,31.85,5.87,15
LR_KSWIN_DATA_cl_F,0.0163,0.0003,2.75,0.51,12
MLPR_none_none_cl_T,0.0164,0,82.53,12.53,41
BRR_KSWIN_MAPE_cl_F,0.0167,0.0008,2.45,0.54,12
MLPR_HDDM_A_DATA_cl_F,0.0169,0,45.36,6.23,24
BRR_KSWIN_DATA_cl_F,0.0171,0.0006,2.71,0.53,12
MLPR_myTanDD_DATA_cl_F,0.019,0,43.9,5.54,21
MLPR_MINPS_MAPE_cl_F,0.0202,0,42.34,5.84,21
MLPR_MINPS_DATA_cl_F,0.0207,0,45.43,7,22
LR_myTanDD_MAPE_cl_F,0.0225,0,3.9,0.43,2
BRR_myTanDD_MAPE_cl_F,0.0232,0,3.82,0.49,2
MLPR_ADWIN_DATA_cl_F,0.0235,0,36.88,5.01,18
MLPR_mySD_DATA_cl_F,0.0242,0,26.14,3.78,11
MLPR_HDDM_W_DATA_cl_F,0.025,0,31.65,4.47,16
MinValInTS_HDDM_A_DATA_cl_F,0.0253,0,0.91,0.31,24
LR_DDM_MAPE_cl_F,0.0269,0,3.45,0.47,1
LR_DDM_DATA_cl_F,0.0269,0,4.58,1.47,1
LR_EDDM_MAPE_cl_F,0.0269,0,3.35,0.34,1
LR_EDDM_DATA_cl_F,0.0269,0,3.5,0.58,1
LR_PH_MAPE_cl_F,0.0269,0,3.42,0.55,1
LR_ADWIN_MAPE_cl_F,0.0269,0,3.59,0.47,1
LR_HDDM_A_MAPE_cl_F,0.0269,0,3.45,0.51,1
LR_HDDM_W_MAPE_cl_F,0.0269,0,3.47,0.56,1
BRR_DDM_MAPE_cl_F,0.0272,0,3.13,0.39,1
BRR_DDM_DATA_cl_F,0.0272,0,3.97,1.26,1
BRR_EDDM_MAPE_cl_F,0.0272,0,3.21,0.48,1
BRR_EDDM_DATA_cl_F,0.0272,0,3.28,0.46,1
BRR_PH_MAPE_cl_F,0.0272,0,3.37,0.55,1
BRR_ADWIN_MAPE_cl_F,0.0272,0,3.39,0.5,1
BRR_HDDM_A_MAPE_cl_F,0.0272,0,3.47,0.56,1
BRR_HDDM_W_MAPE_cl_F,0.0272,0,3.42,0.53,1
MinValInTS_MINPS_DATA_cl_F,0.0304,0,1.12,0.27,22
MLPR_PH_DATA_cl_F,0.032,0,17.21,2.39,7
MinValInTS_MINPS_MAPE_cl_F,0.0325,0,0.99,0.33,23
MinValInTS_mySD_MAPE_cl_F,0.0367,0,1.41,0.3,18
MinValInTS_myTanDD_DATA_cl_F,0.0402,0,0.95,0.24,21
MLPR_KSWIN_DATA_cl_F,0.0405,0.0029,24.9,3.71,12
MLPR_KSWIN_MAPE_cl_F,0.0408,0.0056,24.88,3.28,12
MinValInTS_HDDM_W_DATA_cl_F,0.0458,0,1.22,0.32,16
MinValInTS_mySD_DATA_cl_F,0.0468,0,1.73,0.41,11
MinValInTS_ADWIN_DATA_cl_F,0.0475,0,1.07,0.28,18
MLPR_DDM_MAPE_cl_F,0.0491,0,9.25,1.15,3
MinValInTS_KSWIN_MAPE_cl_F,0.0554,0.0013,1.31,0.37,12
MinValInTS_PH_DATA_cl_F,0.0599,0,1.66,0.37,7
MLPR_ADWIN_MAPE_cl_F,0.0602,0,12.15,2.14,4
MinValInTS_KSWIN_DATA_cl_F,0.0644,0.0047,1.33,0.34,12
MinValInTS_DDM_MAPE_cl_F,0.0867,0,1.93,0.43,4
MLPR_HDDM_A_MAPE_cl_F,0.0994,0,8.19,1.21,2
MinValInTS_ADWIN_MAPE_cl_F,0.11,0,1.74,0.4,5
MLPR_myTanDD_MAPE_cl_F,0.1201,0,8.36,1.33,2
MinValInTS_HDDM_A_MAPE_cl_F,0.1259,0,1.86,0.5,3
MLPR_DDM_DATA_cl_F,0.1413,0,7.46,2.1,1
MLPR_EDDM_MAPE_cl_F,0.1413,0,6.23,1.1,1
MLPR_EDDM_DATA_cl_F,0.1413,0,6.03,0.85,1
MLPR_HDDM_W_MAPE_cl_F,0.1413,0,6.04,0.89,1
MLPR_PH_MAPE_cl_F,0.1445,0,8.29,1.26,2
MinValInTS_myTanDD_MAPE_cl_F,0.1723,0,2.63,0.65,2
MinValInTS_PH_MAPE_cl_F,0.1783,0,1.79,0.44,3
MinValInTS_DDM_DATA_cl_F,0.2356,0,1.93,0.36,1
MinValInTS_EDDM_MAPE_cl_F,0.2356,0,1.84,0.38,1
MinValInTS_EDDM_DATA_cl_F,0.2356,0,1.85,0.33,1
MinValInTS_HDDM_W_MAPE_cl_F,0.2356,0,1.95,0.47,1
\end{filecontents*}
\csvautotabular[respect all]{data/KBEedr.csv}
	\par
	\caption{Time series KBE. Runtime with standard deviation, number of drifts detected with standard deviation, and MAPE for all configurations ordered by the error mean (MAPE). The number of drift point series shows zero standard deviation for all cases.} 
	\label{tabKBE}
\end{table}
Finally, we comment on the MAPE and prediction curves for two specific experiments which have been selected as typical instances of the observed results. Please note then that the curves reported in 
fig. \ref{spypred}, \ref{spyerror},  \ref{kbepred}, and \ref{kbeerror}  are not averaged over several runs but are the results of two single experiments: one made on the SPY time series and the other made on the KBE one. 
Fig. \ref{spypred} and \ref{kbepred} show the real time series and the predicted one. In the case of the YC learner, fig. \ref{kbepred}, an almost precise overlapping occur between the two curves. Whereas in the case of the MinValInTS learner (fig. \ref{spypred}), the predicted curve (the red lines) show the increasing divergence over time between the two. The vertical dashed lines in the figures mark the time when a concept drift has been detected. 
The number of detected drifts depends on the used concept drift detector and on its input values. 
Please note that the Fig. \ref{spyerror} and \ref{kbeerror} report the $MAPE_{last\ k}(t)$ timeseries
introduced in Section \ref{sliding}. 
EAch figure shows how the model learned at time 1, or re-learned after a concept drift, performs over the last k points in the prediction set. 
And, also, show the sensitivity of different concept drift detectors.
In fig. \ref{spyerror}, MAPE reaches the 0.25 value (25\% prediction error) before a concept drift is detected. Whereas in fig. \ref{kbeerror}, MAPE does not increase above 0.06 before a concept drift is detected. 


\section{Computational cost of the proposed methodology \label{compcost} }
Here is an analysis of the computational cost of running one of the learning configurations. 
In the experiments the following parameters have been used:\\

{\small 
\begin{tabular}{l} 
$training\ set\ size = 30 $\\
$cdd_{window\ length} = 20 $\\
$data\_to\_predict  = size(dataset) - training\ set\ size $
\end{tabular} 
}\\

The training set contains 30 instances. In rare case, it could be greater if a c.d.d. 
has marked with a 'warning point', preceding the 'drift point', the possible start of a concept drift.
  
Also some concept drift detectors (c.d.d.) need to collect a set of instances ($cdd_{window}$) before being able to detect a drift point in order to provide statistical validity to their decisions.
All the concept drift detectors from the state of the art have been run with their default parameters.
Instead, our three c.d.d. (myTanDD, mySD, and MINPS) use a $dd_{window}$ of 20 instances.\\

{\small 
\begin{tabular}{l} 
$n\_drifts  =  f(cdd, data\_to\_predict, learner) $
\end{tabular} 
}\\

The number of detected concept drifts, $n\_drifts$, can only be determined in real time because it depends on the incoming data points. 
Therefore, we represent it as a function $f$ of the c.d.d., the $data\_to\_predict$ and the $learner$.

It is however possible to determine a higher bound for $n\_drifts$ if working with a finite subset of a time series: \\
$n\_drifts$ for all configuration $\leq $ the number of drifts detected when using the continuous learning configuration. For instance, in the performed experiments, the number of concept drifts ranges from 1 (no concept drift detected) to 41.\\

{\small 
\begin{tabular}{l} 
$cdd_{ph1}  = n\_drifts *  cdd_{window\ length} $\\
$cdd_{ph2} = data\_to\_predict - cdd_{ph1} $
\end{tabular} 
}\\

Each time a concept drift detector is called, two things may happen either the input data is used to fill up the data detector's window or the input data is used to make a decision about the occurrence of a concept drift. The computational cost of these two activities is different because in the first case the input is simply appended to a list, while in the second case a number of calculations have to be performed. This is why, we decided to keep separate traces of the two activities. 

The cost of filling up the c.d.d.'s window ($cdd_{ph1}$) and the cost to detect a concept drift ($cdd_{ph1}$) are constant for each input value.
 
And finally, the cost of running one learning case, $tm_{ls\_with\_dd}$ is given by: \\ 

{\small 
\begin{tabular}{l} 
$tm_{learn} =  n\_drifts * learning\ cost * training\ set\ size $  \\
$tm_{pred} = data\_to\_predict * prediction\ cost * test\ set\ size $ \\
$tm_{cdd} = cdd_{ph1}* cadd +  cdd_{ph2}* cdd $ \\
$tm_{ls\_with\_dd} = tm_{learn} + tm_{pred} + tm_{cdd} + tm_{updt\_ds} $
\end{tabular} 
}\\

\noindent
where the learning time $tm_{learn}$ and the prediction time $tm_{pred}$ have been assumed linear in the size of the training set. The validity of this assumption depends on the chosen learner. 
Finally, $tm_{updt\_ds}$ represents the cost needed to update the sliding window's data structures. The cost is irrelevant with respect to the other time components.
To provide an example, fig. \ref{mlprcomplexity} shows $tm_{updt\_ds} = t\_i\_perc + t\_rt\_perc$ in the case of the MLPR learner operating on the SPY time series.

\begin{figure*}[thb]
	\centering
	\includegraphics[width=1\linewidth]{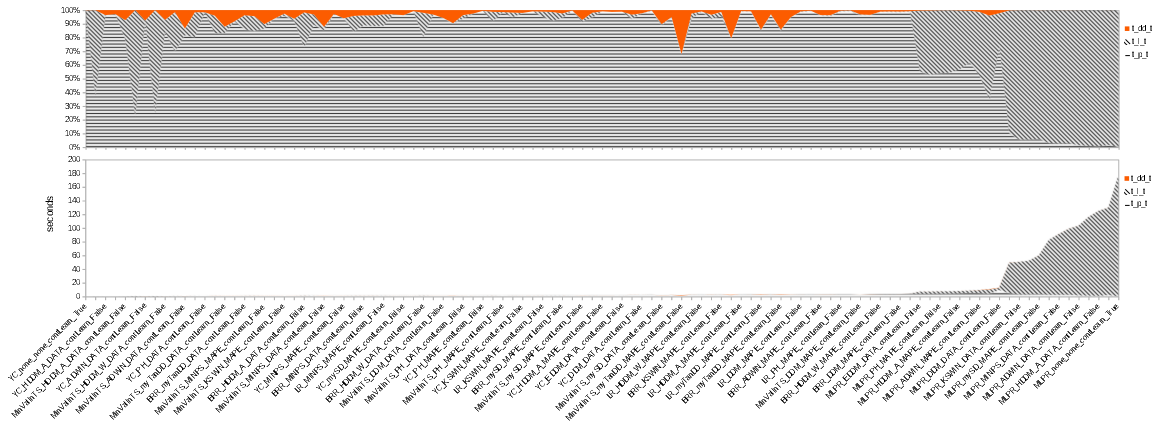}
	\caption{Timeseries SPY - Computational costs of learning, predicting and detecting concept drifts for each experimental configurations in percentage terms (above) and in absolute values. The configuration on the x axis are ordered by increasing runtime.}
	\label{lpdPhases}
\end{figure*}

We provide a graphical representation of the computational costs for the main runtime components in our experiments in fig. \ref{lpdPhases}. 
The figure shows the costs of learning, predicting and detecting concept drifts for each experimental configurations in percentage terms (above) and in absolute values (below). 
Fig. \ref{lpdPhases} show the computational costs in the case of the SPY time series however
 an analogous graph can be produced for the KBE time series. 

\section{Future works and conclusion}
The study investigated the interactions among learners, concept drift detectors, and their input, in the domain of financial time series.
In addition to a better understanding of such interactions, the study recommends two main practical methodologies 
to select promising combinations of the above components when dealing with a new financial time series.

It would not make for an interesting reading the listing here of the several observations made in the experimental section. However, we want to stress that concept drift detectors should be analyzed together with the learning systems they will work with to provide practical insights on their operative strengths and weaknesses.

Following this line of investigation, we aim to extend this research considering the impact that concept drift detectors could have when more algorithmically complex learning systems are used in the domain of financial time series like when deep learning \cite{deeptrading,deeptrading2}, technical analysis \cite{technicaltrading,techtrading2}, reinforcement learning learning \cite{reinforcementtrading,deeptrading2}, evolutionary computation \cite{NeriAIcom12} or agent based simulation \cite{NeriExSyst2020,Neri2019Wivace} are employed.

A second research direction that we aim to pursue is studying if parameter hyper-optimization \cite{Camilleri2014582,Camilleri2014203} can have any impact on a learning system's performance
 when using concept drift detection on financial time series.

A third and final research direction of interest is to analyze more in detail  the working hypothesis discussed in Section \ref{efs} and its possible relationship with the instance representation provided to the learner.

\bibliographystyle{plain}
\bibliography{biblio}

\end{document}